\documentclass[preprint]{aastex63}
\usepackage{CJK}
\usepackage{amsmath,amssymb,amsthm,bm}
\usepackage{graphicx}   
\usepackage{color}      
\usepackage{soul,xcolor}
\usepackage{float}

\begin{document}

\title{Pressure-Strain Interaction as the Energy Dissipation Estimate in Collisionless Plasma}

\author{Yan Yang}
\affiliation{Department of Physics and Astronomy,
University of Delaware, Newark, DE 19716, USA}
\affiliation{Southern University of Science and Technology, Shenzhen, Guangdong 518055, China}
\author{W.H. Matthaeus}
\affiliation{Department of Physics and Astronomy,
University of Delaware, Newark, DE 19716, USA}
\author{Sohom Roy}
\affiliation{Department of Physics and Astronomy,
University of Delaware, Newark, DE 19716, USA}
\author{Vadim Roytershteyn}
\affiliation{Space Science Institute, Boulder, CO 80301, USA}
\author{Tulasi Parashar}
\affiliation{School of Chemical and Physical Sciences, Victoria University of Wellington, Wellington 6012, New Zealand}
\author{Riddhi Bandyopadhyay}
\affiliation{Department of Astrophysical Sciences, Princeton University, Princeton, NJ 08544, USA}
\author{Minping Wan}
\affiliation{Guangdong Provincial Key Laboratory of Turbulence Research and Applications, Department of Mechanics and Aerospace Engineering, Southern University of Science and Technology, Shenzhen, Guangdong 518055, China}

\correspondingauthor{Minping Wan}
\email{wanmp@sustech.edu.cn}

\begin{abstract}
The dissipative mechanism in weakly collisional plasma is a topic that pervades decades of studies without a consensus solution.
We compare several energy dissipation estimates based on energy transfer processes in plasma turbulence and provide justification for the pressure-strain interaction as a direct estimate of the energy dissipation rate. 
The global and scale-by-scale energy balances are examined in 2.5D and 3D kinetic simulations. 
We show that the global internal energy increase and the temperature enhancement of each species are 
directly tracked by the pressure-strain interaction. 
The incompressive part of the pressure-strain interaction dominates over its compressive part in all simulations considered.
The scale-by-scale energy balance is quantified 
by scale filtered Vlasov-Maxwell equations, 
a kinetic plasma approach, 
and the lag dependent von K\'arm\'an-Howarth equation, an approach 
based on fluid models. 
We find that the energy balance is exactly satisfied across all scales, but
the lack of a well-defined inertial range 
influences the distribution of the energy budget among different terms in the inertial range.
Therefore, the widespread use of the Yaglom relation to estimating dissipation rate is questionable in some cases, especially when the scale separation in the system is not clearly defined.
In contrast, the pressure-strain interaction balances exactly the dissipation rate at kinetic scales regardless of the scale separation.
\end{abstract}

\section{Introduction}
The picture of turbulence cascade 
as developed by Taylor, von K\'arm\'an, Kolomogorov and others
\citep{Taylor38,KarmanHowarth38,Kol41a,Kol41b} 
describes energy transfer across scales from 
an energy-containing range, through an inertial range and 
into a dissipation range where fluctuations are converted into internal energy.
This 
basic picture is 
readily adapted to 
magnetohydrodynamics (MHD) \citep{Biskamp-turb}, a magnetofluid model \citep{Moffatt}  often applied as an approximation 
to low density high temperature cosmic plasmas \citep{Parker-cmf} such as the
solar wind \citep{BreechEA08}.
Recently there has been increasing interest
in advancing 
descriptions of turbulence in 
more complex plasma models
\citep{BowersLi07,ParasharEA09,HowesEA11-gyro}.
In particular there 
is impetus to 
understand 
dissipation
for kinetic plasmas 
in which the classical collisional approach becomes inapplicable, and therefore also inapplicable
are standard closures that capture
dissipation as empirical 
descriptions in terms of fluid scale variables.
Fortunately instead of studying specific dissipative mechanisms at kinetic scales,
such as wave-particle interactions \citep{Hollweg86,Hollweg02,Gary03,Gary08,Howes2008kinetic,Markovskii06,He2015evidence,chandran2010perpendicular}, field-particle correlations \citep{klein2016measuring,chen2019evidence,klein2020diagnosing} or heating by coherent structures and reconnection \citep{Dmitruk04,Retino07,Sundkvist07,Parashar11,TenBarge13,Perri12,He2018plasma}, an alternative strategy based on the energy transfer process is available. 
One can proceed by tracing available  pathways
that lead to 
deposition of internal (or thermal) energy. ``Dissipation" here simply refers to a degradation of fluid scale and electromagnetic fluctuations into internal energy \citep{MatthaeusEA20}. This increase in internal
energy is eventually ``thermalized" by collisions and entropy production \citep{PezziEA19-Boltzmann,LiangEA19} but this
irreversibility is not our focus here.

Different dissipation proxies based on the energy transfer process have been adapted to estimate the dissipation rate.
A von K\'arm\'an-Howarth picture \citep{KarmanHowarth38} of turbulent decay was extended for MHD
\citep{hossain1995phenomenology,WanEA12-jfm}, in which the global decay rate of energy is controlled by the von K\'arm\'an decay law at energy-containing scales \citep{WanEA12-jfm,WuEA13-vKH,ZankEA17,BandyopadhyayEA18,BandyopadhyayEA18-prx,Bandyopadhyay2019evolution}. The energy dissipation rate is estimated by the energy decay rate.
The energy loss at energy-containing scales, is then transferred across scales in the MHD inertial range.
To obtain the inertial range 
energy transfer rate, the Yaglom relation \citep{PolitanoPouquet98-grl,Valvo07,Stawarz09,Coburn15,Bandyopadhyay2018solar} and its extensions \citep{Osman11-PRL,Podesta2008laws,Hadid2017energy,AndresEA19,BanerjeeGaltier13-exact,Wan2009third,Hellinger2018karman,Ferrand2019exact,VerdiniEA15} have been adapted, in which the energy dissipation rate is estimated by the energy transfer rate.
The transferred energy ultimately goes into the internal energy reservoir, which is often 
evaluated by the rate of work done by the electric field on particles for species $\alpha$, $\boldsymbol{j}_{\alpha} \cdot \boldsymbol{E}$ \citep{Zenitani2011new,Wan12,Wan15,Chasapis2018situ,Ergun2018magnetic,PhanEA18,lu2019turbulence,voros2019energy,pongkitiwanichakul2021role} and more recently by the pressure-strain interaction for species $\alpha$, $-\left( \boldsymbol{P}_{\alpha} \cdot \nabla \right) \cdot \boldsymbol{u}_{\alpha}$ \citep{YangEA17-PoP,YangEA17-PRE,yang2019energy,Sitnov2018kinetic,Chasapis2018energy,zhong2019energy,BandyopadhyayEA20-PiD,jiang2021statistical}. 

One observable feature is that the aforementioned dissipation rate estimates dominate at different scales. Two standard ways to disentangle the multi-scale properties in configuration space are scale filtering representations and lag dependent structure functions.
Scale filter
\citep{Germano92}
is based on real space coarse-graining
at a specified 
length scale,
which is widely used in Large Eddy Simulation computational
methods \citep{MeneveauKatz00,Pope04},
in analytical theory
\citep{Eyink03,Aluie13},
and recently in promising applications 
to turbulence  
in plasma models
\citep{YangEA17-PoP,Eyink18,YangEA19,TeacaEA21,cerri2020space,Camporeale2018coherent}.
Structure functions such as the von K\'arm\'an-Howarth relation \citep{KarmanHowarth38,MoninYaglom-vol2},
defined in terms of 
increments, have a long history 
in turbulence theory \citep{Frischbook}.
Here we are concerned with energy balance across
scales, measured using the von K\'arm\'an-Howarth equation \citep{VerdiniEA15,Hellinger2018karman,Ferrand2019exact,adhikari2021energy} and
filtered Vlasov-Maxwell equations \citep{YangEA17-PoP,MatthaeusEA20}.
In both 
scale filtering and structure function representations, crucial elements are energy loss at large scales, 
energy flux in an intermediate range, and 
dissipation at small scales. So we compare 
corresponding scale filtered 
and structure function decompositions and clarify the relationship between energy dissipation estimates at different scales.

A number of studies have been devoted to investigate the multiplicity of dissipation rate estimates \citep{YangEA19,MatthaeusEA20,zhou2021measurements,bandyopadhyay2020enhanced,bandyopadhyay2021energy,pezzi2020dissipation}. In this paper we specifically address the role of the pressure-strain interaction for species $\alpha$, $-\left( \boldsymbol{P}_{\alpha} \cdot \nabla \right) \cdot \boldsymbol{u}_{\alpha}$, in two aspects. On the one hand, different pathways contributing to the global evolution of energies are traced, as used in \citep{PezziEA19,song2020force,du2018plasma}, which illustrates the difference between $-\left( \boldsymbol{P}_{\alpha} \cdot \nabla \right) \cdot \boldsymbol{u}_{\alpha}$ and $\boldsymbol{j}_{\alpha} \cdot \boldsymbol{E}$. On the other hand, the scale decomposition of these pathways is realized
using the von K\'arm\'an-Howarth equation and filtered Vlasov-Maxwell equations, which illustrates the relationship among the von K\'arm\'an decay law, the Yaglom relation and $-\left( \boldsymbol{P}_{\alpha} \cdot \nabla \right) \cdot \boldsymbol{u}_{\alpha}$. We make a detailed comparison of these elements using kinetic plasma simulations. Major findings are: {\it first}, that the potential deficiencies of the Yaglom relation to estimating dissipation rate, especially
for the turbulence without well defined inertial range, is exemplified; {\it second},
that energy balance across scales is found, even when the Yaglom relation and its subsidiary assumptions
are not valid; {\it third}, that the 
quantitatively correct description 
of dissipation in kinetic plasma
is that computed from the pressure-strain interaction. 

\section{Theory and Method}
\subsection{Energy Balance Equations}
\label{sec:global-eq}
There are multiple forms of energy available for participation in energy transfer in collisionless plasma. 
These include: electromagnetic energy $E^m=
\left(\boldsymbol{B}^2
+\boldsymbol{E}^2\right)/(8\pi)$, where $\boldsymbol{B}$ and $\boldsymbol{E}$ are magnetic and electric fields;
fluid flow kinetic energy of species $\alpha$ $E^f_\alpha=\rho_\alpha \boldsymbol{u}_\alpha^2/2$, where $\rho_\alpha$ is the mass density and $\boldsymbol{u}_\alpha$ is the fluid velocity;
and 
corresponding internal energy $E^{th}_\alpha={\frac{1}{2}} m_\alpha
\int{\left(\boldsymbol{v}-\boldsymbol{u}_\alpha\right)^2 f_\alpha \left(\boldsymbol{x},\boldsymbol{v},t\right)
d\boldsymbol{v}}$, with mass $m_\alpha$ and velocity distribution function $f_\alpha$.
The first three moments of the Vlasov equation, in
conjunction with the Maxwell equations, yield the time evolution of the energies:
\begin{eqnarray}
\partial_t E^{f}_\alpha + \nabla \cdot \left( E^{f}_\alpha \boldsymbol{u}_\alpha + \boldsymbol{P}_\alpha \cdot \boldsymbol{u}_\alpha \right) &=&
\left( \boldsymbol{P}_\alpha \cdot \nabla \right) \cdot {\boldsymbol{u}}_\alpha+
\boldsymbol{j}_\alpha  \cdot  \boldsymbol{E}, \label{eq:Ef}\\
\partial_t E^{th}_\alpha + \nabla \cdot \left( E^{th}_\alpha \boldsymbol{u}_\alpha + \boldsymbol{h}_\alpha \right) &=& -\left( \boldsymbol{P}_\alpha \cdot \nabla \right) \cdot \boldsymbol{u}_\alpha, \label{eq:Eth}\\
\partial_t E^{m} + {\frac{c}{4\pi}} \nabla \cdot \left( \boldsymbol{E} \times \boldsymbol{B} \right) &=& -\boldsymbol{j} \cdot \boldsymbol{E}, \label{eq:Em}
\end{eqnarray}
where the subscript $\alpha=e, i$ represents the species, $\boldsymbol{P}_\alpha$ is the pressure tensor,
$\boldsymbol{h}_\alpha$ is the heat flux vector,
$\boldsymbol{j}=\sum_{\alpha} \boldsymbol{j}_\alpha$ is the total electric current density,
$\boldsymbol{j}_\alpha=n_\alpha q_\alpha \boldsymbol{u}_\alpha$ is the electric current density of species $\alpha$, and $n_\alpha$ and $q_\alpha$ are the number density and the charge of species $\alpha$, respectively. 
Note that the spatial transport terms (i.e., the second terms on the left hand side) are globally conservative and cancel out under certain boundary conditions, e.g., periodic. 

From these equations, one can see the roles of several energy transfer pathways. For example, the electric
work, $\boldsymbol{j}_{\alpha} \cdot \boldsymbol{E}$, exchanges electromagnetic energy with fluid flow energy for species $\alpha$, while the pressure-strain interaction, $-\left( \boldsymbol{P}_\alpha \cdot \nabla \right) \cdot \boldsymbol{u}_\alpha$, represents the conversion between fluid flow and internal energy for
species $\alpha$.
The pressure-strain interaction 
can be further decomposed as
\begin{equation}
\langle -\left(\boldsymbol{P}_{\alpha} \cdot \nabla\right) \cdot \boldsymbol{u}_{\alpha} \rangle= \langle -p_{\alpha} \nabla \cdot \boldsymbol{u}_{\alpha}-\boldsymbol{\Pi}_{\alpha}:\boldsymbol{D}_{\alpha} \rangle
=p\theta_{\alpha}+PiD_{\alpha},
\label{eq:PS}
\end{equation}
where $\langle \cdots \rangle$ denotes spatial average, $p_\alpha=P_{\alpha,ii}/3$, $\Pi_{\alpha,ij}=P_{\alpha,ij}-p_\alpha \delta_{ij}$, and $D_{\alpha,ij}=\left(\partial_i u_{\alpha,j} + \partial_j u_{\alpha,i}\right)/2-\left(\nabla \cdot \boldsymbol{u}_{\alpha}\right) \delta_{ij}/3$. $p\theta_{\alpha}$ and $PiD_{\alpha}$ denote compressible and incompressible parts respectively.

\subsection{Scale Filtering Representation} 
\label{sec:filtered-eq}
To disentangle the scale-by-scale dynamics, it is useful 
to define a filter \citep{Germano92} at scale
$\ell$, 
operating on the velocity and electromagnetic fields.
With a properly defined filtering kernel $G_\ell\left(\boldsymbol{r}\right)=\ell^{-d}G\left(\boldsymbol{r}/\ell\right)$, where $d$ is the spatial dimension and $G\left(\boldsymbol{r}\right)$ is a non-negative and normalized boxcar window function so that
$\int d^d r G (\boldsymbol{r})=1$,
the field $f(\boldsymbol{x}, t)$  
is filtered at scale $\ell$ as
\begin{equation*}
    \bar{f}_\ell(\boldsymbol{x}, t) = 
\int d^d r G_\ell (\boldsymbol{r}) f(\boldsymbol{x}+\boldsymbol{r}, t).
\end{equation*}
The low-pass filtered $\bar{f}_\ell(\boldsymbol{x}, t)$
only contains information at length scales $\ge\ell$.
The Favre-filtered (density-weighted-filtered) field \citep{Favre69} is defined as 
\begin{equation*}
   \tilde{f}_\ell=\overline{\rho f}_\ell /\bar{\rho}_\ell 
\end{equation*}
We drop the subscript $\ell$ for notation simplicity. So an overbar $\bar{f}$ denotes a filtered quantity
and a tilde $\tilde{f}$ denotes a density-weighted (Favre) filtered quantity. 

Proceeding with the Vlasov-Maxwell equations, 
the resolved kinetic and electromagnetic energy equations at scale $\ell$ \citep{MatthaeusEA20,YangEA17-PoP,Eyink18} are 
\begin{eqnarray}
\partial_t \widetilde{E}^{f}_{\alpha} + \nabla \cdot \boldsymbol{J}^u_{\alpha}
=  - {\Pi}^{uu}_{\alpha}  - {\Phi}^{uT}_{\alpha}
- {\Lambda}^{ub}_{\alpha},\label{eq:filtered-Ef}\\
\partial_t \overline{E}^m + \nabla \cdot \boldsymbol{J}^b =
-\sum_\alpha {\Pi}^{bb}_{\alpha}
+\sum_\alpha {\Lambda}^{ub}_{\alpha}, \label{eq:filtered-Em}
\end{eqnarray}
where $\widetilde{E}^{f}_{\alpha} =\frac{1}{2} \bar{\rho}_{\alpha} \tilde{\boldsymbol{u}}_{\alpha}^2$ is the filtered fluid flow energy; 
$\overline{E}^m = (\bar{\boldsymbol{B}}^2+\bar{\boldsymbol{E}}^2)/(8\pi)$ is the filtered electromagnetic energy; 
$\boldsymbol{J}^u_{\alpha}={\widetilde{E}^{f}_{\alpha} \tilde{\boldsymbol{u}}_{\alpha} + \bar{\rho}_{\alpha} \tilde{\boldsymbol{\tau}}^{u}_{\alpha} \cdot \tilde{\boldsymbol{u}}_{\alpha} + \overline{\boldsymbol{P}}_{\alpha} \cdot \tilde{\boldsymbol{u}}_{\alpha}}$ and $\boldsymbol{J}^b=\left(c \overline{\boldsymbol{E}} \times \overline{\boldsymbol{B}}\right)/(4\pi)$ are the spatial transport;
$\Pi^{uu}_{\alpha}=-\left(\bar{\rho}_{\alpha} \tilde{\boldsymbol{\tau}}^{u}_{\alpha} \cdot \nabla\right) \cdot \tilde{\boldsymbol{u}}_{\alpha} - q_{\alpha}/c \bar{n}_{\alpha} \tilde{\boldsymbol{\tau}}^{b}_{\alpha} \cdot \tilde{\boldsymbol{u}}_{\alpha}$ is the sub-grid-scale (SGS) flux of fluid flow energy across scales due to nonlinearities, where $\tilde{\boldsymbol{\tau}}^{u}_{\alpha} = \widetilde{\boldsymbol{u}_{\alpha}\boldsymbol{u}_{\alpha}}-\tilde{\boldsymbol{u}}_{\alpha}\tilde{\boldsymbol{u}}_{\alpha}$, $\tilde{\boldsymbol{\tau}}^{b}_{\alpha} = \widetilde{\boldsymbol{u}_{\alpha} \times \boldsymbol{B}}-\tilde{\boldsymbol{u}}_{\alpha}\times \tilde{\boldsymbol{B}}$;
$\Pi^{bb}_{\alpha}=-q_\alpha \bar{n}_\alpha  \tilde{\boldsymbol{\tau}}^{e}_{\alpha} \cdot \tilde{\boldsymbol{u}}_\alpha$ is the as the SGS flux of electromagnetic energy across scales due to nonlinearities, where $\tilde{\boldsymbol{\tau}}^{e}_{\alpha} = \tilde{\boldsymbol{E}}-\bar{\boldsymbol{E}}$;
$\Phi^{uT}_{\alpha}=-\left(\overline{\boldsymbol{P}}_{\alpha} \cdot \nabla\right) \cdot \tilde{\boldsymbol{u}}_{\alpha}$ is the rate of flow energy converted into internal energy through filtered $-\left(\boldsymbol{P}_{\alpha} \cdot \nabla\right) \cdot \boldsymbol{u}_{\alpha}$;
$\Lambda^{ub}_{\alpha}=-q_{\alpha} \bar{n}_{\alpha} \widetilde{\boldsymbol{E}} \cdot \tilde{\boldsymbol{u}}_{\alpha}$ is the rate of conversion of fluid flow energy into electromagnetic energy through filtered $-\boldsymbol{j}_{\alpha}\cdot \boldsymbol{E}$.

We can avoid the complication of the interchange between fluid flow and electromagnetic energy and the spatial transport by summing Eqs. \ref{eq:filtered-Ef} and \ref{eq:filtered-Em} over species and averaging over space, which yields
\begin{equation}
\underbrace{\partial_t \left<\sum_\alpha \widetilde{E}^{f}_{\alpha} + \overline{E}^m\right>}_{T_f-\epsilon}
=-\underbrace{\left<\sum_\alpha \left( {\Pi}^{uu}_{\alpha} + {\Pi}^{bb}_{\alpha}\right)\right>}_{F_f}- \underbrace{\left<\sum_\alpha {\Phi}^{uT}_{\alpha}\right>}_{D_f} \label{eq:filtered-Ef+Em}.
\end{equation}
The spatial transport terms are globally conservative under suitable
boundary conditions, e.g., periodic. $\epsilon$ is the total dissipation rate. In the absence of an exact expression of dissipation in collisionless plasma simulations, the dissipation rate is computed by $\epsilon=-d\left<\sum_\alpha E^{f}_{\alpha} +E^m\right>/dt$.  
The first term is the time-rate-of-change of energy at scales $\ge\ell$, which is $0$ at large enough $\ell$ and $-\epsilon$ at $\ell \rightarrow 0$. In analogy with the structure function representation in Sec. \ref{sec:sf-eq}, the first term is written as $T_f=\epsilon + \partial_t \left<\sum_\alpha \widetilde{E}^{f}_{\alpha} + \overline{E}^m\right>$, which is the time-rate-of-change of energy at scales $<\ell$ and equals to $\epsilon$ at large enough $\ell$ (e.g., larger than energy-containing scales) and 0 at $\ell \rightarrow 0$. The remaining terms are the nonlinear cross-scale 
energy flux $F_f$ and the deposition of 
internal energy received from the cascade $D_f$.
Similar to Eq. \ref{eq:PS}, the filtered pressure-strain interaction 
is decomposed as
\begin{equation}
\langle \Phi^{uT}_{\alpha}\rangle=\langle -\left(\overline{\boldsymbol{P}}_{\alpha} \cdot \nabla\right) \cdot \tilde{\boldsymbol{u}}_{\alpha}\rangle=\langle-\bar{p}_{\alpha} \nabla \cdot \tilde{\boldsymbol{u}}_{\alpha}\rangle-\langle\overline{\boldsymbol{\Pi}}_{\alpha}:\widetilde{\boldsymbol{D}}_{\alpha}\rangle=\overline{p\theta}_{\alpha}+\overline{PiD}_{\alpha}.
\label{eq:filtered-PS}
\end{equation}

\subsection{Structure Function Representation}
\label{sec:sf-eq}
The energy distribution among scales can also be defined by considering two-point correlations or the related second-order structure functions. We proceed with the incompressible Hall MHD model due to 
its relative analytical simplicity. The associated 
von K\'arm\'an-Howarth equation \citep{KarmanHowarth38,MoninYaglom-vol2,VerdiniEA15,Hellinger2018karman,Ferrand2019exact,adhikari2021energy}
in structure function form is
\begin{equation}
    \underbrace{\partial_t S(\boldsymbol{\ell})/4}_{-T_s} + \underbrace{\nabla_{\boldsymbol{\ell}} \cdot \left[\boldsymbol{Y}(\boldsymbol{\ell})/4+\boldsymbol{H}(\boldsymbol{\ell})/8\right]}_{-F_s} = -\epsilon + \underbrace{D_{\mu}(\boldsymbol{\ell}) + D_{\eta}(\boldsymbol{\ell})}_{D_s} \label{eq:SF},
\end{equation}
where $\boldsymbol{\ell}$ is the spatial lag, $\nabla_{\boldsymbol{\ell}}$ is the gradient with respect to the lag $\boldsymbol{\ell}$, $S(\boldsymbol{\ell})=\langle \rho \left(\delta \boldsymbol{u}\right)^2 \rangle+\langle \left(\delta \boldsymbol{B} \right)^2 \rangle$ is a second-order structure function, $\boldsymbol{u}$ the bulk fluid velocity, and $\delta \boldsymbol{u}=\boldsymbol{u}(\boldsymbol{x}+\boldsymbol{\ell})-\boldsymbol{u }(\boldsymbol{x})$ defines the increment.
$\boldsymbol{Y}(\boldsymbol{\ell})=\langle \left(\delta \boldsymbol{u}\right)^2 \delta \boldsymbol{u} + \left(\delta \boldsymbol{B} \right)^2 \delta \boldsymbol{u} -2\left(\delta \boldsymbol{B} \cdot \delta \boldsymbol{u}\right) \delta \boldsymbol{B}\rangle$ is a third-order structure function, $\boldsymbol{H}(\boldsymbol{\ell})=\langle 2\left(\delta \boldsymbol{B} \cdot \delta \boldsymbol{j}\right) \delta \boldsymbol{B} - \left(\delta \boldsymbol{B} \right)^2 \delta \boldsymbol{j}\rangle$ is the Hall term, and 
$D_{\mu}(\boldsymbol{\ell})=\mu \nabla_{\boldsymbol{\ell}}^2 \langle \left(\delta \boldsymbol{u}\right)^2 \rangle/2$
and $D_{\eta}(\boldsymbol{\ell})=\eta \nabla_{\boldsymbol{\ell}}^2 \langle \left(\delta \boldsymbol{B} \right)^2 \rangle/2$ are the viscous and resistive terms, respectively. 

Several points concerning Eq. \ref{eq:SF} warrant clarification. 
First, the terms in Eq. \ref{eq:SF} are functions of lag vector $(\ell_x, \ell_y, \ell_z)$. Upon averaging over directions, the terms in Eq. \ref{eq:SF} only depend on lag length $\ell$ and $T_s$, $F_s$ and $D_s$ denote omnidirectional forms of the three terms in Eq. \ref{eq:SF}.
Second, Eq. \ref{eq:SF} refers to a fluid model, so $\boldsymbol{u} = (m_i n_i \boldsymbol{u}_i +m_e n_e \boldsymbol{u}_e)/(m_i n_i +m_e n_e)$ and
$\rho = m_i n_i +m_e n_e$ are dominated by contributions from ions.
Third, for the incompressible model used here, we neglect the spatial and time variability of density and set it to be a constant that equals to the spatially and temporally averaged density. The kinetic plasma simulations used here are weakly compressed, which are shown in detail in Sec. \ref{sec:global-result} and Sec. \ref{sec:scale-result}. The incompressible model therefore remains a credible approximation for our simulations.
For
strongly compressed plasmas, e.g., 
the turbulent magnetosheath, more elaborate compressive
models \citep{BanerjeeGaltier13-exact,Hadid2017energy,AndresEA19} are required.
Next, the interpretation of
$S/4$ is that it is related 
to the energy at scales $<\ell$. To see this,
let the lag $\ell \rightarrow 0$, so that $S/4$ tends to zero, while it equals to the fluid flow and magnetic energy at large $\ell$. 
The second term  $F_s$ measures the energy flux through the surface of a lag sphere of radius $\ell$. 
Finally, the viscous and resistive term $D_s$ has the property that for incompressible MHD, it converges to the dissipation rate $\epsilon=\mu\langle \nabla \boldsymbol{u} : \nabla \boldsymbol{u}\rangle + \eta \langle \nabla \boldsymbol{b} : \nabla \boldsymbol{b}\rangle$ in the limit $\ell\rightarrow 0$.
Since this study proceeds with kinetic plasma simulations,
we cannot compute $\epsilon$ and $D_s$ directly, which instead are derived by $\epsilon=-d\left<\sum_\alpha E^{f}_{\alpha} +E^m\right>/dt$ and $D_s=\epsilon-T_s-F_s$.

\subsection{Association with Dissipation Rate Estimates}
To appreciate the physical content of the comparisons below, it is necessary to understand both the 
differences and the similarities of the energy balance Equations \ref{eq:Ef}-\ref{eq:Em}, the 
scale filtering formulation Eq. \ref{eq:filtered-Ef+Em} and the structure function 
formulation Eq. \ref{eq:SF}, and their associations with dissipation rate.
The energy balance equations show direct causality between the global evolutions of energies and $\boldsymbol{j}_{\alpha} \cdot \boldsymbol{E}$ and $-\left( \boldsymbol{P}_\alpha \cdot \nabla \right) \cdot \boldsymbol{u}_\alpha$, while the
scale filtering and structure function 
formulations show details of energy distributions and dissipation proxies across scales.

The scale filter as implemented in Eq. \ref{eq:filtered-Ef+Em} contains the scale-decomposed energy budget of the full 
Vlasov Maxwell model. 
This includes compressive and wave particle effects and contributions from both protons and electrons.
The structure function model as implemented 
in Eq. \ref{eq:SF}
is a purely fluid construct (incompressible Hall MHD).
As such it lacks
compressible effects, wave-particle interactions, and other non-Hall kinetic effects.
Therefore, we anticipate that the correspondence  between Eqs. \ref{eq:filtered-Ef+Em} and \ref{eq:SF}
may remains credible at MHD scales, identified tentatively as
scales larger than ion inertial length.

Even still, the scale-decomposed formulations
share common elements in the physics they represent. Both are expressions of 
conservation of energy across scales and are composed of
(i) the rate of change of energy at scales $<\ell$, $T_f$ and $T_s$;
(ii) energy transfer
across scale $\ell$, $F_f$ and $F_s$, leading to the possibility of an inertial range; and (iii) measures of energy dissipation, $D_f$ and $D_s$.
These elements are in direct association with energy dissipation proxies based on the energy transfer process. For example, the time derivative terms $T_f$ and $T_s$ at energy-containing $\ell$ represent the energy decay rate, which can be taken as a dissipation estimate. According to the von K\'arm\'an decay law \citep{WanEA12-jfm,WuEA13-vKH,ZankEA17,BandyopadhyayEA18,Bandyopadhyay2019evolution}, 
\begin{equation}
    \epsilon_{vK,\pm}=-\frac{d(Z^{\pm})^2/2}{dt}=\frac{\alpha_{\pm}}{2} \frac{(Z^{\pm})^2 Z^{\mp}}{L_{\pm}},
    \label{eq:vk}
\end{equation}
where $\alpha_{\pm}$ are positive constants, $Z^{\pm}$ are the rms fluctuation values of the Elsasser variables, and $L_{\pm}$ are 
similarity length scales (correlation lengths are usually used),
the energy decay rate is given by $\epsilon_{vK}=(\epsilon_{vK,+}+\epsilon_{vK,-})/2$.
The terms $F_f$ and $F_s$ are related to the inertial range 
energy transfer rate, giving rise to the widely used Yaglom relation \citep{PolitanoPouquet98-grl,Valvo07,Stawarz09,Coburn15,Bandyopadhyay2018solar,VerdiniEA15}
\begin{equation}
    \epsilon_{Ym}=-\nabla_{\boldsymbol{\ell}} \cdot \left(\frac{\boldsymbol{Y}}{4}+\frac{\boldsymbol{H}}{8}\right),
    \label{eq:ym}
\end{equation}
which can taken as a dissipation estimate.
Although the exact expression of $D_s$ is not known in our kinetic plasma simulations, the term $D_f$ shown in Eq. \ref{eq:filtered-Ef+Em} is the spatial average of filtered pressure-strain interaction, which can estimate the fluid flow energy conversion into internal energy.

In summary, the intent of energy balance equations in Sec. \ref{sec:global-eq} is to illuminate global behaviors, while the scale filtering and structure function representations in Sec. \ref{sec:filtered-eq} and Sec. \ref{sec:sf-eq} aim at scale dependence. Detailed treatments based on the three aforementioned formulations are useful to describe the extent to which these dissipation proxies are correlated and their domain of validity. 

\section{Kinetic Simulations: 2.5D and 3D}
Vlasov-Maxwell 
solutions are obtained with particle in cell (PIC) 
codes;
for run parameters, see
Table \ref{tab:2.5d-3d-params}.
Here we make use of 2.5 dimensional (2.5D) and 3D kinetic simulations with no external drive. So both are decaying initial value problems.

The 2.5D case was performed
using the code P3D \citep{Zeiler02} in a 2.5D geometry
with turbulent fluctuations in $(x, y)$ plane but no variation in the
$z$ direction. All physical quantities have all three components of
the vectors.
The simulation was performed in a periodic domain, whose size
is $L\simeq 150 d_i$, where $d_i$ is the ion inertial length, with $4096^2$ grid points and
$3200$ particles of each species
per cell ($\sim 1.07\times 10^{11}$ total particles).
The ion-to-electron mass ratio is $m_i/m_e = 25$,
and the ratio $\omega_{pe}/\omega_{ce}=3$, where $\omega_{pe}$ is the electron plasma frequency and $\omega_{ce}$ is the electron cyclotron frequency.
The run was started with uniform density $n_0=1.0$
and Maxwellian-distributed ions and electrons with temperature $T_0=0.3$.
The uniform magnetic field is $B_0 = 1.0$ directed
out of the plane, and plasma $\beta=0.6$.
The velocity and magnetic perturbations are transverse to $B_0$, typical of ``Alfv\'enic" modes.
They were seeded at time $t=0$ by
populating Fourier modes for a range of wavenumbers $2\le |\boldsymbol{k}| \le 4$
with random phased fluctuations and a specific spectrum. 
The run was evolved for more than $300 \omega_{ci}^{-1}$. This simulation was also used in \citep{ParasharApJL18,YangEA19,MatthaeusEA20,bandyopadhyay2021energy}.

The 3D case was obtained using the VPIC code \citep{Bowers08}. The simulation was conducted in a fully periodic 3D domain of size $L\simeq 42 d_i$ with resolution of $2048^3$ cells and $150$ particles per cell per species ($\sim 2.6\times 10^{12}$ total particles). The ion-to-electron mass ratio is $m_i/m_e=50$ and the ratio $\omega_{pe}/\omega_{ce}=2$.
The initial conditions correspond to uniform plasma with density $n_0$,
Maxwellian-distributed ions and electrons of equal temperature $T_0$, uniform magnetic field $B_0$ in $z$ direction, and plasma $\beta=0.5$. Turbulent fluctuations were seeded at time $t=0$ by imposing a large-scale isotropic spectrum of velocity and magnetic fluctuations having polarizations transverse to the guide magnetic field $B_0$. The run was evolved for about $170 \omega_{ci}^{-1}$. Further details of this simulation can be found in \citep{RoytershteynEA14}. 

\begin{table}[H]
    \centering
    \begin{tabular}{cccccccc}
    \hline
	    Dimension  & $L(d_i)$ & $N$ & $m_i/m_e$ & $B_0 \hat{z}$  & $\delta b/B_0$ & $\beta$ & $ppg$\\
     \hline
	    2.5D    & $150$ & 4096 & 25 & 1.0 & 0.5 & 0.6 & 3200 \\
     \hline
          3D    & $42$ & 2048 & 50 & 0.5 & 1.0 & 0.5 & 150 \\
     \hline
    \end{tabular}
    \caption{2.5D and 3D PIC simulation parameters: domain size $L$; grid points in each direction $N$; 
ion-to-electron mass ratio $m_i/m_e$; guide magnetic field in  z-direction $B_0$; initial magnetic fluctuation amplitude $\delta b$; plasma $\beta$; average number of particles of each species per grid $ppg$.}
    \label{tab:2.5d-3d-params}
\end{table}

\section{Global Behavior of Pressure-Strain Interaction}
\label{sec:global-result}
We study the time evolution of global volume averages of energy to
find the electromagnetic work $\langle \boldsymbol{j}_{\alpha} \cdot \boldsymbol{E} \rangle$ versus the pressure-strain interaction $\langle -\left( \boldsymbol{P}_\alpha \cdot \nabla \right) \cdot \boldsymbol{u}_\alpha \rangle$, the incompressive  $PiD_{\alpha}$ versus compressive $p\theta_{\alpha}$ conversion of energy, and ion ($\alpha=i$) versus electron ($\alpha=e$) heating. Accurate integrals of energy transfer pathways over time require small time steps and this time-integrated procedure is rather expensive for the 3D PIC simulation. Therefore, the analysis in this section is only conducted using the 2.5D PIC simulation.

\begin{figure}[H]
    \centering
    \includegraphics[width=0.75\textwidth]{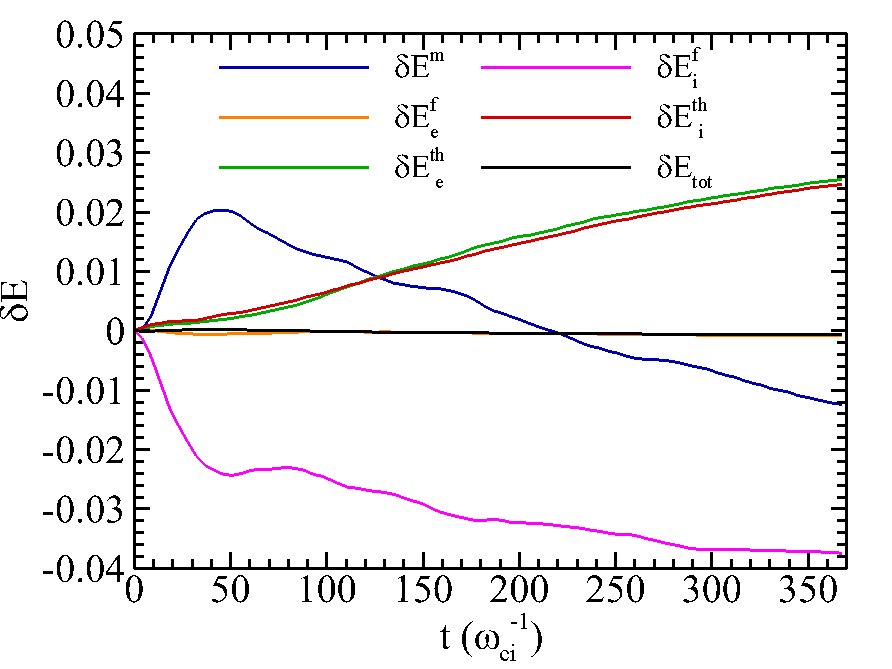}
    \caption{Time evolution of the electromagnetic energy $\langle E^m \rangle$, the fluid flow energy of each species $\langle E^f_{\alpha}\rangle$, and the internal energy of each species $\langle E^{th}_{\alpha}\rangle$ for the 2.5D simulation. The change of energy is defined as $\delta E(t)=E(t)-E(0)$. The change of the total energy $\langle E_{tot}\rangle= \langle E^m + E^f_i + E^f_e + E^{th}_i + E^{th}_e\rangle$ is also shown to verify excellent energy conservation.}
    \label{fig:2d-evo-energy}
\end{figure}

Fig. \ref{fig:2d-evo-energy} shows the global energy balance by tracking time evolution of the electromagnetic energy $\langle E^m \rangle$, the fluid flow energy of each species $\langle E^f_{\alpha}\rangle$, and the internal energy of each species $\langle E^{th}_{\alpha}\rangle$. The total energy $\langle E_{tot}\rangle= \langle E^m + E^f_i + E^f_e + E^{th}_i + E^{th}_e\rangle$ is well conserved, indicating the validity of this simulation. Note that the electrons gain slightly more internal energy compared to the ions. The stronger electron energization is in conflict with the findings of \citep{CranmerEA09,BandyopadhyayEA20-PiD,hughes2014electron} which favor stronger ion heating, while this is consistent with what is observed in \citep{bandyopadhyay2021energy} who found stronger electron heating in turbulent magnetic reconnection. 
As suggested in early studies, the partitioning of heating between ions and electrons depends on turbulence amplitude \citep{WuEA13-vKH,MatthaeusEA16,hughes2017kinetic} and plasma $\beta$ \citep{howes2010prescription,ParasharApJL18,vech2017nature}.

\begin{figure}[H]
    \centering
    \includegraphics[width=0.49\textwidth]{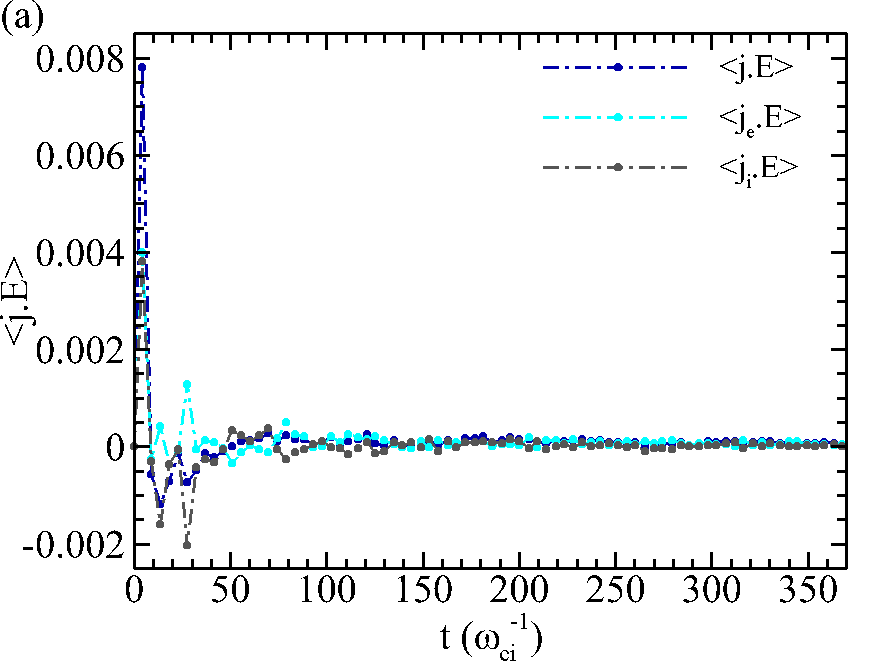}
    \includegraphics[width=0.49\textwidth]{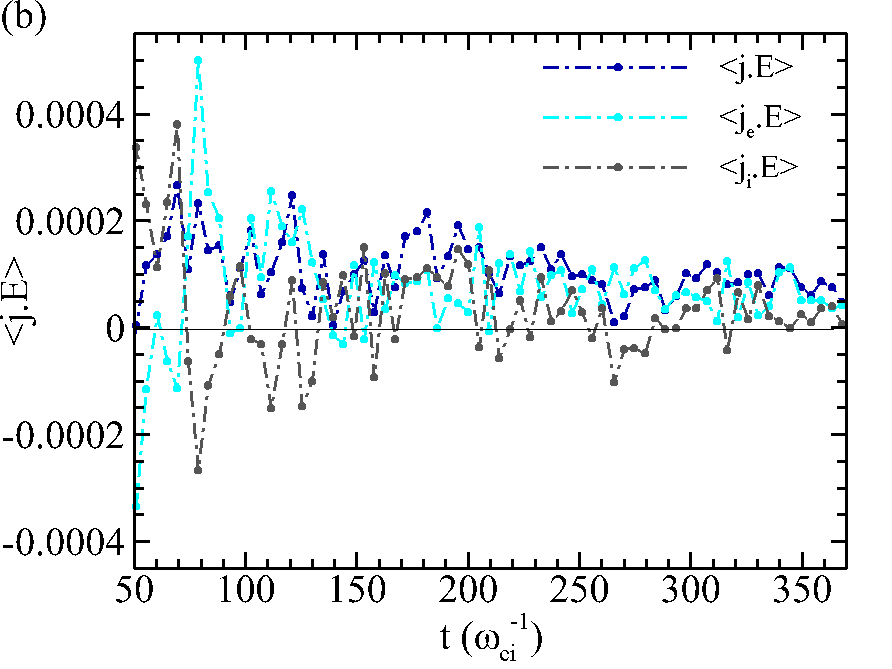}
    \includegraphics[width=0.49\textwidth]{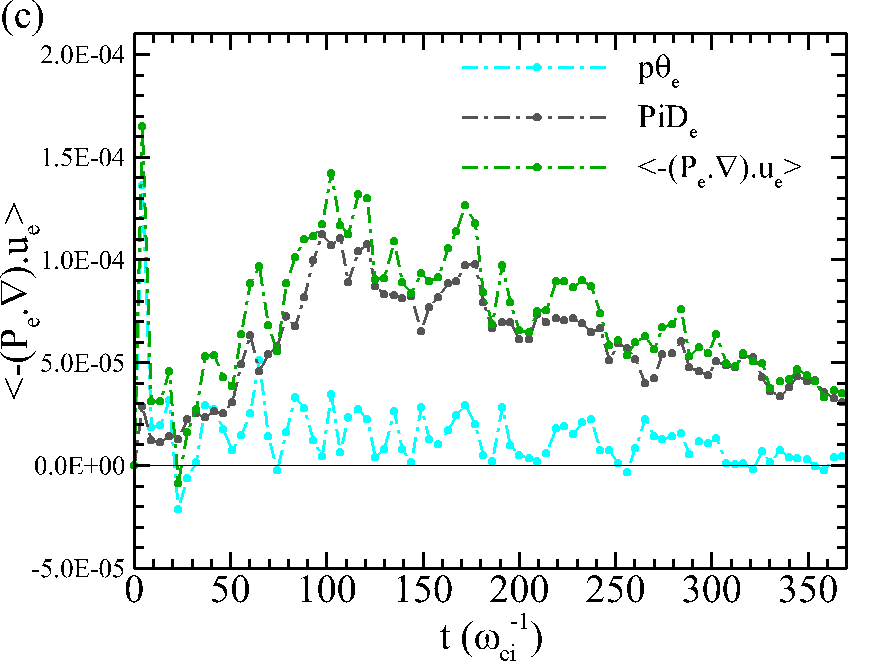}
    \includegraphics[width=0.49\textwidth]{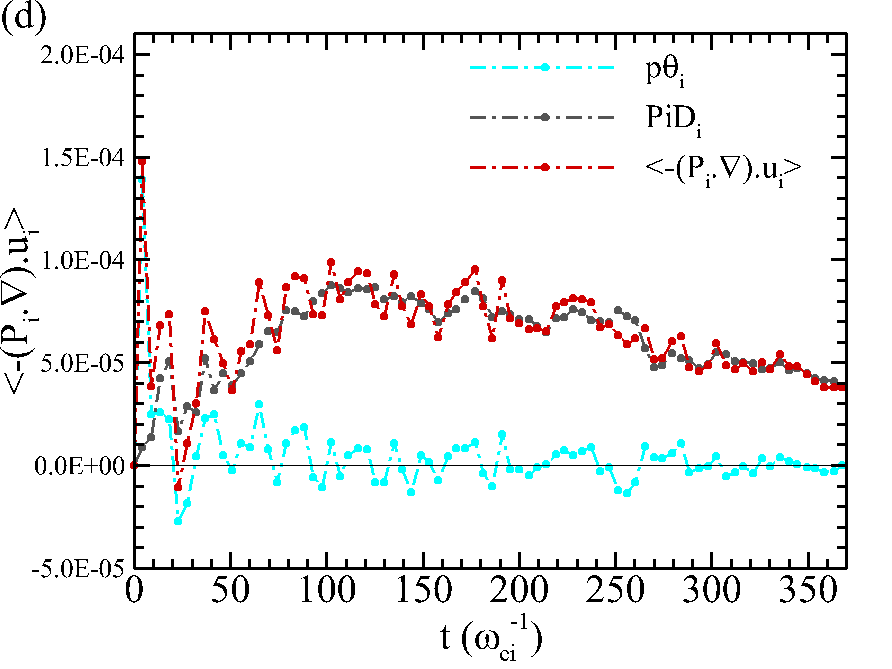}
    \caption{Time evolution of (a) $\langle \boldsymbol{j}_i  \cdot  \boldsymbol{E} \rangle $, $\langle \boldsymbol{j}_e  \cdot  \boldsymbol{E} \rangle$ and $\langle \boldsymbol{j}  \cdot  \boldsymbol{E} \rangle$;
    (b) zoomed-in electric work showing oscillations;
    (c) $p\theta_e$, $PiD_e$, and $\langle -\left( \boldsymbol{P}_e \cdot \nabla \right) \cdot \boldsymbol{u}_e \rangle$; 
    and (d) $p\theta_i$, $PiD_i$, and $\langle -\left( \boldsymbol{P}_i \cdot \nabla \right) \cdot \boldsymbol{u}_i \rangle$ for the 2.5D simulation.}
    \label{fig:2d-evo-JE-PS}
\end{figure}

From Eqs. \ref{eq:Ef}-\ref{eq:Em}, two pathways, the electric work $\langle \boldsymbol{j}_{\alpha} \cdot \boldsymbol{E} \rangle$ and the pressure-strain interaction $\langle -\left( \boldsymbol{P}_\alpha \cdot \nabla \right) \cdot \boldsymbol{u}_\alpha \rangle$, contribute to the global energy exchange between forms. Their time histories are shown in Fig. \ref{fig:2d-evo-JE-PS}. One can observe that the global average of the electric work of each species $\langle \boldsymbol{j}_{\alpha} \cdot \boldsymbol{E} \rangle$ in Fig. \ref{fig:2d-evo-JE-PS}(a) and (b) oscillates significantly over time at high frequencies. {This is likely an artefact of the artificial value of $\omega_{pe}/\omega_{ce}$ in our simulation,
and may be remedied by time averaging the results over a plasma oscillation period \citep{HaggertyEA17}.} The global average of the pressure dilatation of each species $p\theta_{\alpha}$ in Fig. \ref{fig:2d-evo-JE-PS} (c) and (d) also exhibits oscillations, which is attributable to the compression arising from acoustic waves. Since the run we use here is weakly compressible, the incompressive channel $PiD_{\alpha}$ is favored obviously over the compressive channel $p\theta_{\alpha}$ for both ions and electrons. Therefore, the global average of the pressure-strain interaction of each species 
$\langle -\left( \boldsymbol{P}_\alpha \cdot \nabla \right) \cdot \boldsymbol{u}_\alpha \rangle$
mainly results from the incompressive part $PiD_{\alpha}$. However, we cannot rule out the possibility of stronger $p\theta_{\alpha}$ at some locations. For example, as shown in the magnetosheath \citep{Chasapis2018energy}, the turbulent current sheet \citep{pezzi2020dissipation,bandyopadhyay2021energy} and the electron diffusion region \citep{zhou2021measurements}, the local compressive part could dominate over the local incompressible part. Note that there is one peak with magnitude much larger than others in the beginning at $t=4.5 \omega_{ci}^{-1}$. This is due to the initial negligible electric field which responds quickly to obey a solution of the Vlasov-Maxwell system. 

\begin{figure}[H]
    \centering
    \includegraphics[width=1.0\textwidth]{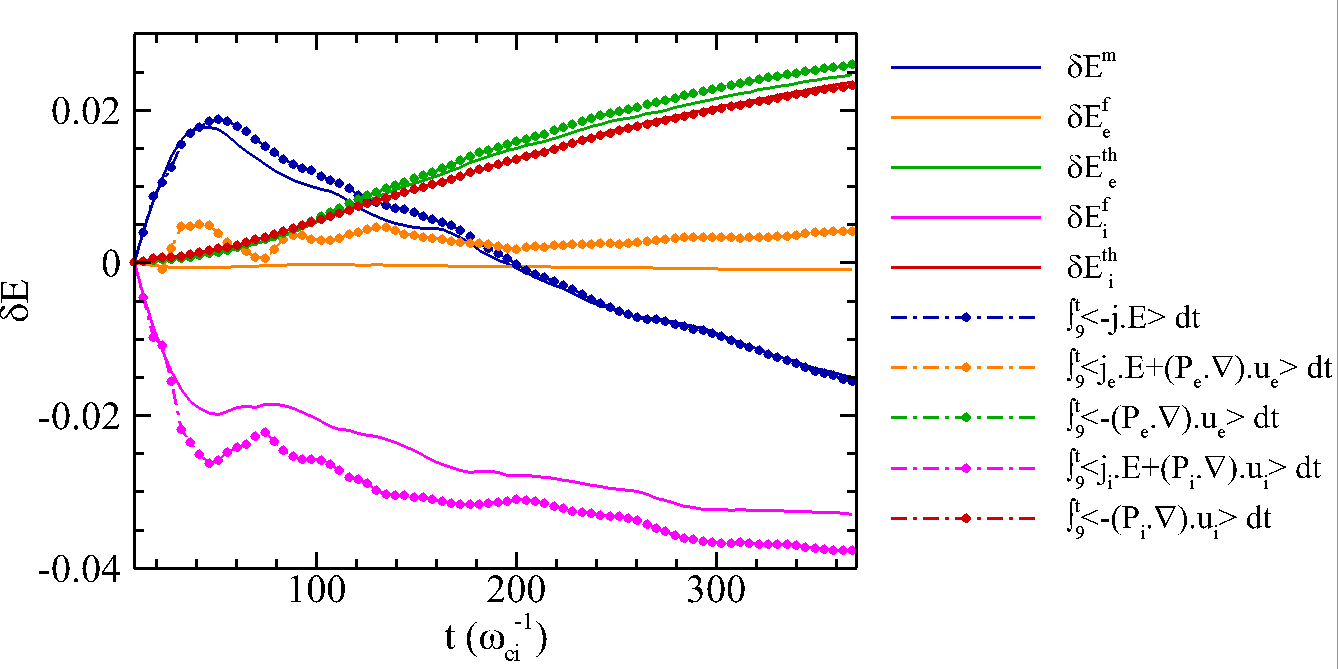}
    \caption{Time evolution of the changes of energies versus cumulative time-integrated 
    $\langle -\boldsymbol{j}  \cdot  \boldsymbol{E} \rangle$, 
    $\langle -\left( \boldsymbol{P}_e \cdot \nabla \right) \cdot \boldsymbol{u}_e \rangle$, 
    $\langle -\left( \boldsymbol{P}_i \cdot \nabla \right) \cdot \boldsymbol{u}_i \rangle$, 
    $\langle \boldsymbol{j}_e  \cdot  \boldsymbol{E} +  \left( \boldsymbol{P}_e \cdot \nabla \right) \cdot \boldsymbol{u}_e \rangle $, and 
    $\langle \boldsymbol{j}_i  \cdot  \boldsymbol{E} + \left( \boldsymbol{P}_i \cdot \nabla \right) \cdot \boldsymbol{u}_i\rangle$ for the 2.5D simulation.
    Here the change of energy is defined as $\delta E(t)=E(t)-E(9 \omega_{ci}^{-1})$ and the cumulative integral is computed over time $[9 \omega_{ci}^{-1}, ~t]$.}
    \label{fig:2d-evo-energy-JE-PS}
\end{figure}

Global energy balance in Eqs. \ref{eq:Ef}-\ref{eq:Em} is shown in Fig. \ref{fig:2d-evo-energy-JE-PS}, where the changes of energies contrast with the cumulative time-integrated pathways. These integrals have been numerically computed through the trapezoidal rule and over time inteval $[9 \omega_{ci}^{-1}, ~t]$ to avoid the effect of the peaks at $t=4.5 \omega_{ci}^{-1}$ shown in Fig. \ref{fig:2d-evo-JE-PS}. According to Eq. \ref{eq:Eth}, the internal energy variation of species $\alpha$, $\delta E^{th}_{\alpha}$, should be equal to the cumulative integral of $\langle -\left( \boldsymbol{P}_\alpha \cdot \nabla \right) \cdot \boldsymbol{u}_\alpha \rangle$, which is confirmed in Fig. \ref{fig:2d-evo-energy-JE-PS}. One can see that the cumulative time integrated $\langle -\left( \boldsymbol{P}_\alpha \cdot \nabla \right) \cdot \boldsymbol{u}_\alpha \rangle$ is almost superposed on the change of internal energy for both electrons and ions. The slight difference mainly results from the level of accuracy of the energy conservation recovered in the simulation and the numerical error of time integration. 

As expected from Eq. \ref{eq:Em}, the cumulative time integrated $\langle -\boldsymbol{j}  \cdot  \boldsymbol{E} \rangle$ is in good agreement with the change of the electromagnetic energy, $\delta E^m$. This adds evidence to the idea that despite being adopted widely to estimate heating rate, the electric work $\langle \boldsymbol{j}  \cdot  \boldsymbol{E} \rangle$ is not in direct association with either the internal energy increase or the temperature enhancement. Instead, the change of the internal energy takes place directly through the pressure-strain interaction, $\langle -\left( \boldsymbol{P}_\alpha \cdot \nabla \right) \cdot \boldsymbol{u}_\alpha \rangle$, for both ions and electrons. 

Further examining Fig. \ref{fig:2d-evo-energy-JE-PS}, we 
see that 
the change of the fluid flow energy $\delta E^f_{\alpha}$ shows similar trends to the cumulative $\langle \boldsymbol{j}_{\alpha}  \cdot  \boldsymbol{E} +  \left( \boldsymbol{P}_{\alpha} \cdot \nabla \right) \cdot \boldsymbol{u}_{\alpha} \rangle $, 
in accord with Eq. \ref{eq:Ef}.
The non-negligible 
offsets of the two curves 
arise in large part from the 
cumulative numerical error 
associated 
with the 
high frequency oscillations
in 
$\langle \boldsymbol{j}_{\alpha}  \cdot  \boldsymbol{E} \rangle$ shown in Fig. \ref{fig:2d-evo-JE-PS}.

\section{Energy Balance across Scales}
\label{sec:scale-result}
The preceding section investigates global properties of energy conversion
in detail.
Another important property 
of plasma turbulence 
is that it encompasses a vast range of scales. Therefore
justification for the identification of relevant 
dissipation proxies at different scales is crucial.
Two simple but essential approaches to resolve or decompose 
turbulent fields over 
varying scales are 
the scale filtering technique,
and structure function technique, 
which give rise to Eq. \ref{eq:filtered-Ef+Em} and Eq. \ref{eq:SF}, respectively. 

To proceed numerically using 
Eq. \ref{eq:SF}, 
the term $\partial_t S(\boldsymbol{\ell})/4$ 
is computed by
$\left(S(\boldsymbol{\ell},t+\Delta t)-S(\boldsymbol{\ell},t)\right)/\Delta t$. 
The gradient $\nabla_{\boldsymbol{\ell}}$ is computed in lag $(\ell_x, \ell_y)$ space spanning
$[0,75d_i]\times[0,75d_i]$ with $256^2$ mesh points in the 2.5D case, 
and in lag $(\ell_x, \ell_y, \ell_z)$ space spanning
$[0,21d_i]\times[0,21d_i]\times[0,21d_i]$ with $64^3$ mesh points in the 3D case. 
To extract the lag length dependence, we apply directional averaging to obtain
lag-length-dependent versions of the 
quantities $T_s$, $F_s$ and $D_s$.

\begin{figure}[H]
    \centering
    \includegraphics[width=0.49\columnwidth]{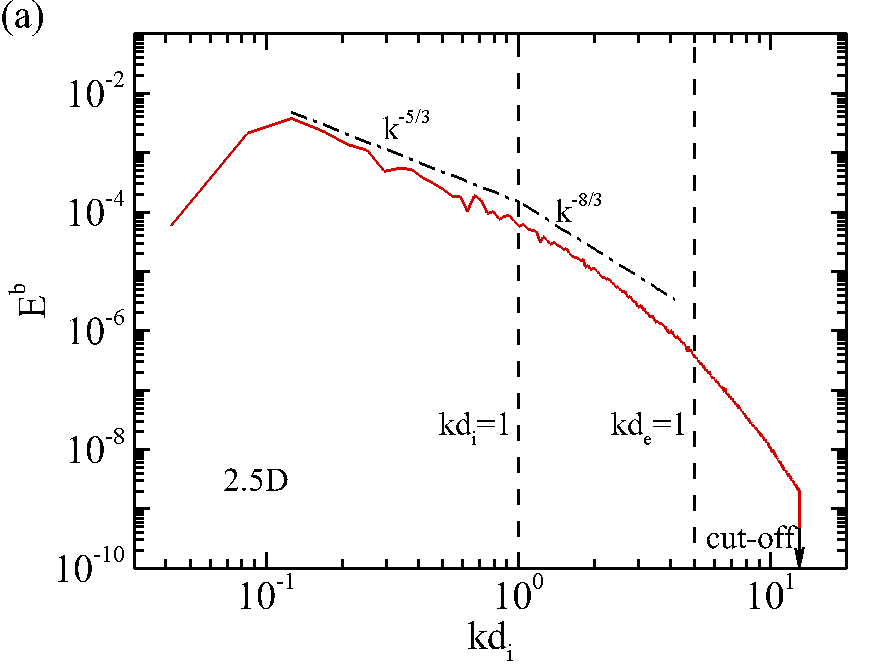}
    \includegraphics[width=0.49\columnwidth]{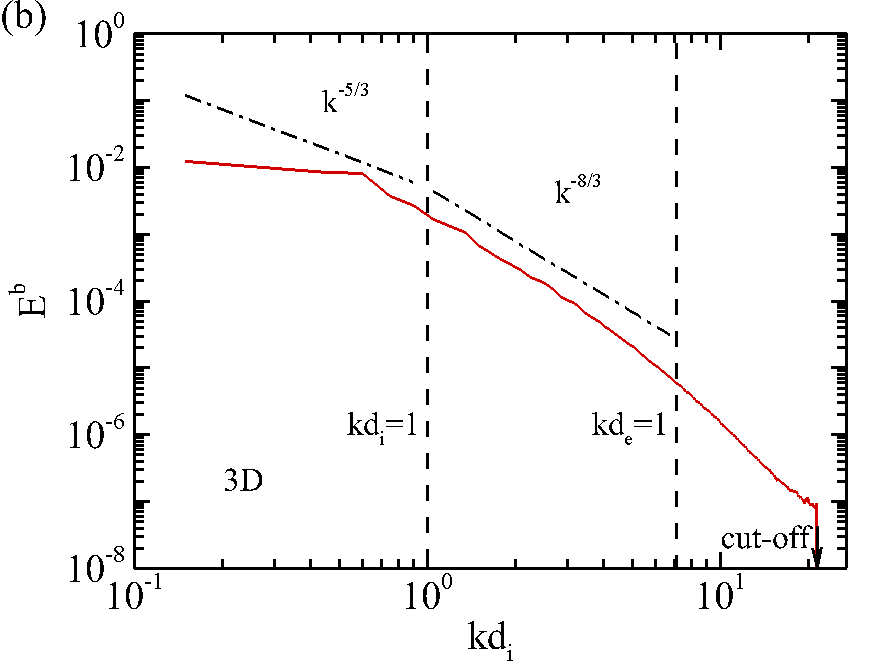}
    \caption{Omnidirectional energy spectra of magnetic fluctuations in 2.5D ($t=97.5\omega_{ci}^{-1}$) 
    and 3D ($t=45\omega_{ci}^{-1}$) PIC simulations.
    Power laws are shown for reference. Vertical lines correspond to ion and electron inertial scales. 
    Filter cut-off is indicated by arrows, as explained in the text.}
    \label{fig:2d-3d-spectra}
\end{figure}

The analysis procedure
selects a time shortly after the mean square current density reaches its maximum, by which time the  
turbulence is fully established.
We use 4 time snapshots (i.e., $t=97.5\omega_{ci}^{-1},~102.5\omega_{ci}^{-1},~107\omega_{ci}^{-1},~111.5\omega_{ci}^{-1}$) in the 2.5D case and 2 time snapshots (i.e., $t=45\omega_{ci}^{-1},~50\omega_{ci}^{-1}$) in the 3D case. These snapshots are separated by time lag $\Delta t \sim 5\omega_{ci}^{-1}$.
All following results are averaged over these time snapshots.
Fig. \ref{fig:2d-3d-spectra} shows magnetic energy 
spectra for the 2.5D case at $t=97.5\omega_{ci}^{-1}$ 
and the 3D case at $t=45\omega_{ci}^{-1}$, 
where $k^{-5/3}$ and $k^{-8/3}$ power laws are shown for reference in the range $k<1/d_i$ and $1/d_i<k<1/d_e$, respectively. 

In comparison to the 2.5D simulation, the setup of the 3D kinetic simulation is 
less than optimal, for example in its relatively small 
domain size. 
The expectation follows that for  
the 3D case the interval of MHD inertial scales will be more limited and 
not well separated from the injection scale.
The spectra are characterized with an upturn at high wavenumbers, which is a cumulative effect of the discrete particle noise inherent in the PIC algorithm. Prior to analysis, we remove noise by
low-pass Fourier filter with a cut-off at $kd_i \sim 13$ for the 2.5D case and $kd_i \sim 21$ for the 3D case. This 
corresponds to filter scales $\ell=\pi/k\sim 0.25 d_i$ and $\ell\sim 0.15 d_i$, respectively. 
This filtering procedure has a negligible
effect on the following results at scales larger than the cut-off. 

\begin{figure}[H]
    \centering
    \includegraphics[width=0.75\columnwidth]{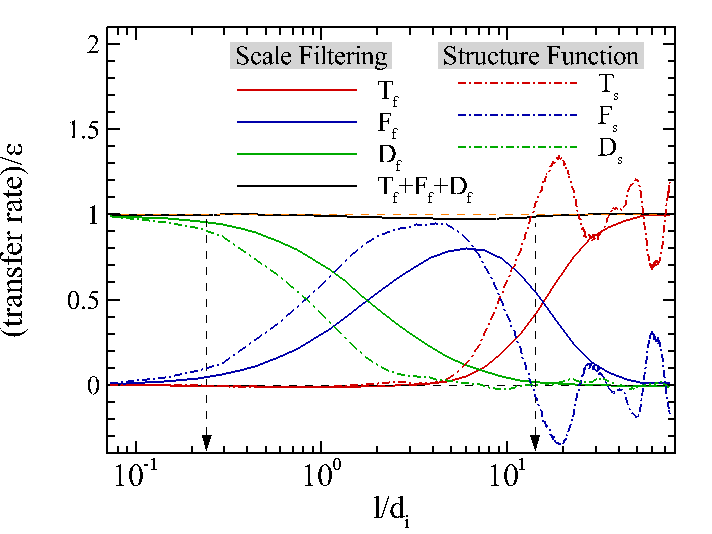}
    \caption{2.5D PIC result: terms in energy budget equations in terms of scale filters (Eq. \ref{eq:filtered-Ef+Em}) and structure functions (Eq. \ref{eq:SF}) with varying scales. 
    The dissipation rate is computed by $\epsilon=-d\left<\sum_\alpha E^{f}_{\alpha} +E^m\right>/dt$. Vertical dashed lines with arrows indicate filter cut-off and correlation length.}
   \label{fig:2d-filter-vs-sf}
\end{figure}

Analysis of the 2.5D case 
begins with a scale-filtering
description of energy balance in Eq. \ref{eq:filtered-Ef+Em}, shown in Fig. \ref{fig:2d-filter-vs-sf}.
The filter cut-off is $\ell= 0.25d_i\sim d_e$ and the correlation length is $\ell\sim 14d_i$, which are indicated by arrows in Fig. \ref{fig:2d-filter-vs-sf}. The $k^{-5/3}$ power law spectrum in Fig. \ref{fig:2d-3d-spectra}(a) gives a rough suggestion of an MHD inertial range, in the vicinity
of a few $d_i$. 
One can see that the time derivative term
$T_f$, which is the time rate of change of the cumulative energy in fluctuations at scales $<\ell$,
reaches the total dissipation rate $\epsilon$ at scales larger than the correlation length,
and decreases at intermediate scales (roughly the MHD inertial range), 
where the nonlinear transfer term $F_f$ sets in and reaches a peak value.
The plateau (or, peak) 
of $F_f$ deviates slightly from $\epsilon$ due to the onset of pressure-strain interaction $D_f$, which is not negligible at the scale of peak $F_f$, and which 
increases towards $\epsilon$
at scales much smaller than $d_i$.

A similar balance holds for the structure function representation in Eq. \ref{eq:SF}. 
In this case, the dissipative term $D_s$,
important at small scales, 
is computed through $D_{\mu} + D_{\eta}=\epsilon+\partial_t S/4+ \nabla_{\boldsymbol{r}} \cdot \left(\boldsymbol{Y}/4+\boldsymbol{H}/8\right)$.
Again, the Yaglom flux $F_s$, representing energy transfer by turbulence, 
reaches a plateau over the intermediate (inertial) scales, at values approaching
the system dissipation rate $\epsilon$. 
The time derivative term $T_s$ saturates to $\sim \epsilon$ at very large scales.

From Fig. \ref{fig:2d-filter-vs-sf}, we
observe the following. First, the scale-by-scale energy budget equation in terms of structure functions (Eq. \ref{eq:SF}), 
although shifted slightly, is in good agreement with the analogous expressions in terms of scale filters (Eq. \ref{eq:filtered-Ef+Em}). Second, the analogy that exists between the filtered pressure-strain interaction $D_f$ and the 
dissipative term $D_s$ lends credence to the association 
of the  pressure-strain interaction with the dissipation rate.
Finally, the energy loss at energy-containing scales equals the nonlinear energy flux at MHD inertial range and the energy dissipated at kinetic scales. Such a well-distinguished range of scales is indicative of a well-separated inertial range. 
Accordingly, for cases such as this one, 
the energy decay rate estimated by the von K\'arm\'an decay law
in Eq. \ref{eq:vk} and the inertial range energy rate estimated by the Yaglom relation in Eq. \ref{eq:ym}
are reliable proxies of energy dissipation rate.

\begin{figure}[H]
    \centering
    \includegraphics[width=0.75\columnwidth]{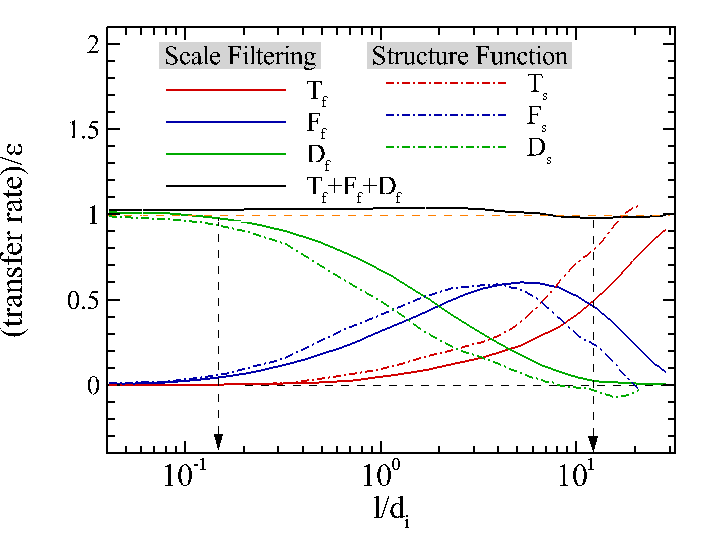}
    \caption{3D PIC result: the same terms as those in the 2.5D PIC simulation shown in Fig. \ref{fig:2d-filter-vs-sf} with varying scales.}
   \label{fig:3d-filter-vs-sf}
\end{figure}

All systems do not necessarily realize a well-defined scale separation as 
seen in the 2.5D case shown above, see for example the observational result in \citep{BandyopadhyayEA20-Hall,ChhiberEA18}. 
We use the 3D PIC run, in which scale separation is less well defined, 
to show how the balance of terms appears 
in Fig. \ref{fig:3d-filter-vs-sf}.
Most of points made in Fig. \ref{fig:2d-filter-vs-sf} remain applicable in Fig. \ref{fig:3d-filter-vs-sf}, such as the analogy between the scale-filtering and structure function representations, and the similarity between pressure-strain interaction and viscous-resistive effect.
In the 3D case (Fig. \ref{fig:3d-filter-vs-sf}),
the dissipation rate is well accounted for 
by the pressure-strain interaction at kinetic scales 
and by the time-rate-of-change of filtered 
energy at energy-containing scales. 
However, both the nonlinear transfer term $F_f$ and the Yaglom flux $F_s$ saturate to values significantly 
below $\epsilon$. This is evidently 
due to the fact that 
 neither the dissipation $D_f$ ($D_s$) nor the time dependence 
 $T_f$ ($T_s$) are negligible at those scales. 

To 
obtain a ``pure''
Yaglom relation (Eq. \ref{eq:ym}), 
one requires 
the existence of the MHD inertial range, in which the dynamics is dominated by inertia terms.
It is therefore necessary that the MHD inertial range of scales is
(i) at scales small enough that 
the time-rate-of-change of energy at that scale is negligible, and (ii) at scales 
large enough 
that the dissipation at that scale can be neglected. 
In this sense, the 3D case shown in Fig. \ref{fig:3d-filter-vs-sf} manifestly 
lacks a well-defined scale separation or a well-separated MHD inertial range.
This is consistent with the spectrum in Fig. \ref{fig:2d-3d-spectra}(b), but is actually more dramatically illustrated in the analysis shown in the scale decomposition shown in Fig. \ref{fig:3d-filter-vs-sf}. 
Consequently, in this 3D case
the estimation of dissipation rate based on the Yaglom relation (Eq. \ref{eq:ym}) 
falls well below the correct one. 
Unlike the Yaglom relation, the pressure-strain interaction term traces the conversion of fluid flow energy into internal energy directly, 
and provides an accurate 
dissipation rate estimation even in the absence of a well-defined 
MHD inertial range, as seen in Figs. \ref{fig:2d-filter-vs-sf} and \ref{fig:3d-filter-vs-sf}.

\begin{figure}[H]
    \centering
    \includegraphics[width=0.49\textwidth]{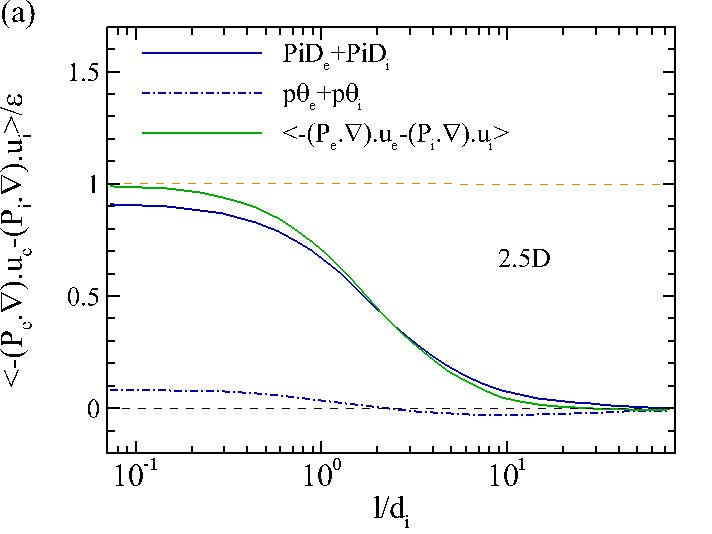}
    \includegraphics[width=0.49\textwidth]{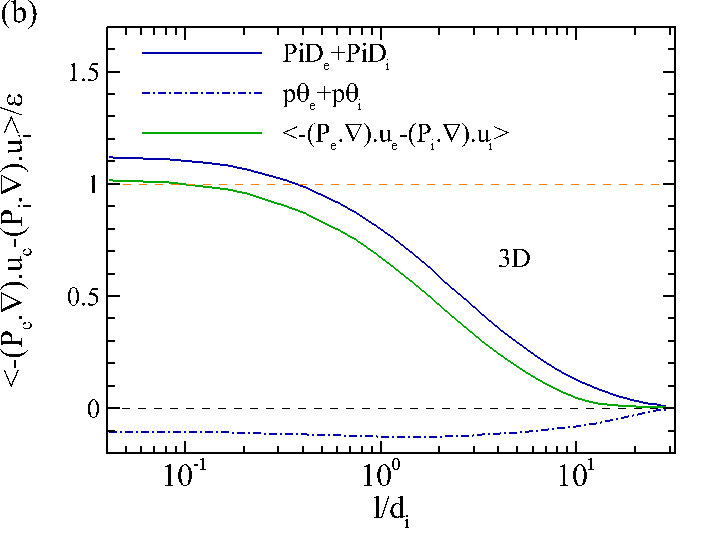}
    \caption{Compressible and incompressible parts in the filtered pressure-strain interaction (Eq. \ref{eq:filtered-PS}) for the 2.5D and 3D PIC simulations.}
    \label{fig:2d-3d-PS}
\end{figure}

As shown in Fig. \ref{fig:2d-evo-JE-PS}, the compressible
part $p\theta_{\alpha}$ accounts for a very small fraction of the pressure-strain interaction 
$\langle -\left( \boldsymbol{P}_\alpha \cdot \nabla \right) \cdot \boldsymbol{u}_\alpha \rangle$ for ions and electrons.
In Fig. \ref{fig:2d-3d-PS}, we show how the compressible $\overline{p\theta}_i+\overline{p\theta}_e$ and 
incompressible $\overline{PiD}_i+\overline{PiD}_e$ parts in Eq. \ref{eq:filtered-PS} vary with scales.
One can see that the compressible part across scales saturates at kinetic scales, but it is much smaller than the incompressible part for both 2.5D and 3D cases. The compressible part $\overline{p\theta}_i+\overline{p\theta}_e$ for the 3D case is negative, while that for the 2.5D case is positive. One explanation for this sign difference is the oscillatory character of $p\theta_{\alpha}$ 
seen in Fig. \ref{fig:2d-evo-JE-PS} due to the involvement of acoustic waves.

\section{Conclusion}
The dissipative mechanism in weakly collisional plasma 
is a topic that pervades decades of studies without a
consensus solution. 
A number of dissipation proxies emerge in the turbulence
energy transfer process.
In this paper, we study the pressure-strain interaction $-\left( \boldsymbol{P}_\alpha \cdot \nabla \right) \cdot \boldsymbol{u}_\alpha$ versus other proxies, such as the electromagnetic
work $\boldsymbol{j} \cdot \boldsymbol{E}$, the energy decay rate $\epsilon_{vk}$ from the von K\'arm\'an decay law
\citep{WanEA12-jfm,WuEA13-vKH,ZankEA17,BandyopadhyayEA18,Bandyopadhyay2019evolution}
and the energy transfer rate $\epsilon_{Ym}$ from the Yaglom relation
\citep{PolitanoPouquet98-grl,Valvo07,Stawarz09,Coburn15,Bandyopadhyay2018solar,VerdiniEA15}.
Both the global energy balance and energy budget across scales using the scale filtering and structure function representations
\footnote{The guide magnetic field $B_0$ does not appear explicitly in the scale filtering and structure function representations \citep{WanEA12-jfm}.
We expect that both Eqs. \ref{eq:filtered-Ef+Em} and \ref{eq:SF} remain applicable in 2D and 3D and even in the presence of guide magnetic field,
which is confirmed by checking the energy balance in our 2.5D and 3D kinetic simulations.}
are investigated in detail in 2.5D and 3D kinetic simulations. 

We confirm that although the electromagnetic 
work $\boldsymbol{j} \cdot \boldsymbol{E}$ has been widely used, the change of the internal energy of each species take places directly through the pressure-strain interaction
$-\left( \boldsymbol{P}_\alpha \cdot \nabla \right) \cdot \boldsymbol{u}_\alpha$.
In comparison with the electric work $\boldsymbol{j} \cdot \boldsymbol{E}$, the pressure-strain interaction $-\left( \boldsymbol{P}_\alpha \cdot \nabla \right) \cdot \boldsymbol{u}_\alpha$
has the advantage of tracking electron and ion heating separately. 
Meanwhile, there can 
be a correlation between the electromagnetic 
work and the pressure-strain interaction. For example, 
from a generalized Ohm's law or the electron momentum equation,
in the limit of massless electrons, we can find that 
$\langle \boldsymbol{j}_e \cdot \boldsymbol{E} \rangle=\langle -\left( \boldsymbol{P}_e \cdot \nabla \right) \cdot \boldsymbol{u}_e \rangle$.  No such relation can be obtained for protons. 

The detailed treatment of energy transfer across scales 
using the scale filtering and structure function representations 
demonstrates significant caveats present in typical estimations of dissipation based on the von K\'arm\'an decay law and the Yaglom relation, while the same analysis
shows the 
the advantages of the pressure-strain interaction $\langle -\left( \boldsymbol{P}_\alpha \cdot \nabla \right) \cdot \boldsymbol{u}_\alpha \rangle$ over other proxies to estimating energy dissipation rate.
At energy-containing scales, the time-rate-of-change of energy can be taken as the dissipation rate. However, in real data environment from observations, 
due to the infeasibility of obtaining the time derivative term,
we resort to the von K\'arm\'an decay law (Eq. \ref{eq:vk})  to compute the energy decay rate. This may be inaccurate due to 
uncertainty of the choices for the similarity lengths $L_{\pm}$ and the von K\'arm\'an constants $\alpha_{\pm}$.

In the MHD inertial range of scales, 
existing studies of energy dissipation in kinetic plasmas 
have had a particular leaning toward use of the Yaglom relation in observations and simulations.
The Yaglom flux 
$\epsilon_{Ym}$ in Eq. \ref{eq:ym} is frequently computed 
or approximated,
with only rare attempts made to circumscribe its domain of validity. 
While we are aware of the important recent developments in deriving Yaglom-like relations for compressible MHD, we have not opted in the present study to implement these theories due to their general complexity, variety of forms and specialization (in some cases) to isothermal turbulence.
The incompressive form that we use here evidently remains 
relevant to our weakly compressible kinetic simulations.
Our 2.5D kinetic simulation exhibits a range of scales (inertial range) 
over which the inertial range energy transfer rate from the Yaglom relation 
fits well with the real dissipation rate,
while observably underestimated inertial range energy transfer rate 
emerges in our 3D case.
This is attributable to the lack of a well-separated inertial range in the 3D kinetic simulation used here.
It is worth emphasizing that
the Yaglom relation requires that there exists a well-separated inertial range, but well-defined scale
separation is not guaranteed a priori. Consequently the prediction that follows
from the Yaglom relation should be questionable in the absence of a well-defined inertial range.

In contrast, the pressure-strain interaction $\langle -\left( \boldsymbol{P}_\alpha \cdot \nabla \right) \cdot \boldsymbol{u}_\alpha \rangle$ turns out to be an accurate estimation of the real dissipation rate even in the absence of a well-defined inertial range.
The pressure-strain interaction dominates at kinetic scales $d_e$, so that deep in the dissipation range, all dissipation is accounted for.
One may note that the pressure-strain interaction
is also analogous to the viscous-collisional case
in the sense that its global average could depend on velocity gradient as well
\citep{Vasquez12,DelsartoPegoraro18,YangEA17-PoP,YangEA-PRE-17,ParasharApJL18}.

In principle the pressure-strain interaction 
$\langle -\left( \boldsymbol{P}_\alpha \cdot \nabla \right) \cdot \boldsymbol{u}_\alpha \rangle$
is a much more complete representation of dissipation in kinetic plasma in several ways. Firstly, $\langle -\left( \boldsymbol{P}_\alpha \cdot \nabla \right) \cdot \boldsymbol{u}_\alpha \rangle$ is readily resolved into contributions of electrons and ions (and additional species if present). Therefore this approach is useful to understand the distribution of dissipated energy between different species \citep{Sitnov2018kinetic,bandyopadhyay2021energy}. 
Secondly, $\langle -\left( \boldsymbol{P}_\alpha \cdot \nabla \right) \cdot \boldsymbol{u}_\alpha \rangle$
includes both compressive and incompressive contributions,
while most of other estimates are limited to incompressive contributions.
The kinetic simulations used here are weakly compressive, and more samples with stronger compression are required to assess detailed contributions from the compressive part of $\langle -\left( \boldsymbol{P}_\alpha \cdot \nabla \right) \cdot \boldsymbol{u}_\alpha \rangle$ to the energy dissipation \citep{Chasapis2018energy,du2018plasma,pezzi2020dissipation,WangEA21-MMS,zhou2021measurements,bandyopadhyay2021energy}.
Thirdly,  $\langle -\left( \boldsymbol{P}_\alpha \cdot \nabla \right) \cdot \boldsymbol{u}_\alpha \rangle$ remains applicable to anistropic cases, 
in which we can figure out the preferential dissipation direction \citep{song2020force}. 
Finally, without taking spatial average, $-\left( \boldsymbol{P}_\alpha \cdot \nabla \right) \cdot \boldsymbol{u}_\alpha$
is spatially localized, which 
can be applied to examine the contribution from each point in space and identify sites of heating \citep{YangEA-PoP-17,YangEA-PRE-17,YangEA19,PezziEA19}. We should also keep in mind that the transport terms shown in Eq. \ref{eq:Eth} 
are likely 
to exert significant influence on the 
values of $-\left( \boldsymbol{P}_\alpha \cdot \nabla \right) \cdot \boldsymbol{u}_\alpha$ \citep{du2020energy,fadanelli2021energy} actually obtained, and may therefore influence
dissipation even if the pressure-strain remains an accurate 
quantitative measure of the dissipation itself.



\acknowledgments 
This research has been supported by NSFC Grant Nos. 91752201, 11672123 and 11902138, by  the 
US National Science Foundation under NSF-DOE grant PHYS- 2108834, by the NASA Magnetospheric Multiscale mission under NASA grant 80NSSC19K0565, and by NSF/DOE grant DE-SC0019315 and XSEDE allocation TG-ATM180015. We acknowledge computing support provided by Center for Computational Science and Engineering of Southern University of Science and Technology
and the use of Information Technologies (IT) resources at the University of Delaware, specifically the high-performance computing resources.

 \newcommand{\BIBand} {and} 
  \newcommand{\boldVol}[1] {\textbf{#1}} 
  \providecommand{\SortNoop}[1]{} 
  \providecommand{\sortnoop}[1]{} 
  \newcommand{\stereo} {\emph{{S}{T}{E}{R}{E}{O}}} 
  \newcommand{\au} {{A}{U}\ } 
  \newcommand{\AU} {{A}{U}\ } 
  \newcommand{\MHD} {{M}{H}{D}\ } 
  \newcommand{\mhd} {{M}{H}{D}\ } 
  \newcommand{\RMHD} {{R}{M}{H}{D}\ } 
  \newcommand{\rmhd} {{R}{M}{H}{D}\ } 
  \newcommand{\wkb} {{W}{K}{B}\ } 
  \newcommand{\alfven} {{A}lfv{\'e}n\ } 
  \newcommand{\alfvenic} {{A}lfv{\'e}nic\ } 
  \newcommand{\Alfven} {{A}lfv{\'e}n\ } 
  \newcommand{\Alfvenic} {{A}lfv{\'e}nic\ }

\end{document}